\documentclass[aps,pra,twocolumn,showpacs]{revtex4}
\usepackage{latexsym,amssymb,amsmath,graphicx}

\begin{document}

\title{Quantum noise radar: superresolution with quantum antennas by accessing spatiotemporal correlations}

\author{I. Peshko$^{1,2}$, D. Mogilevtsev$^{2}$, I. Karuseichyk$^{2}$, A. Mikhalychev$^{2}$, A. P. Nizovtsev$^{2}$, G. Ya. Slepyan$^{3}$, A. Boag$^{3}$}

\affiliation{$^{1}$Physics Department, Belarusian State University, Bobruyskaya st. 5, Minsk  220030, Belarus; \\
$^{2}$Institute of Physics, National Academy of Sciences of Belarus, Nezavisimosti Ave. 68, Minsk 220072, Belarus; \\
$^{3}$School of Electrical Engineering, Tel Aviv University, Tel Aviv 69978, Israel}




\begin{abstract}
We suggest overcoming the "Rayleigh catastrophe" and reaching superresolution for imaging with both spatially and temporally-correlated field of a superradiant quantum antenna. Considering far-field radiation of two interacting spontaneously emitting two-level systems, we show that for the measurement of the temporally-delayed second-order correlation function of the scattered field, the Fisher information does not tend to zero with diminishing the distance between a pair of scatterers even for non-sharp time-averaged detection. For position estimation of a larger number of scatterers, measurement of the time-delayed function is able to provide a considerable accuracy gain over the zero-delayed function. We show also that the superresolution with the considered quantum antenna can be achieved for both near-field imaging and estimating parameters of the antenna.
\end{abstract}


\maketitle

\section{Introduction}

Radar is a system that uses electromagnetic waves (pulsed or continuous) to obtain information about the object ("target") by detecting the field scattered by this object. The simplest task for which radar has been used is target detection or, in another words, decision whether the object is present or absent inside a given solid angle. Another type of the problem is to decide whether the object is stationary or moves with respect to the radar, and what the value of its velocity is. The maximally complicated task is object configuration recovery (imaging) via the analysis of its scattering pattern in a lossy and noisy environment  \cite{richards2010principles}.

 For classical imaging, there are well-known limits for the resolutions defined by the wavelength of the imaging field and parameters of the imaging set-up \cite{classic}. When the details of the imaged object become much smaller than the limiting value, a possibility to discern these details is lost. Statistically, it appears like a necessity to carry out measurement for unrealistically long time for reaching accuracy sufficient for discerning the details in question \cite{helstrom1970resolvability}.  This phenomenon can be aptly described as loss of the information and formalized with help of information measures, such as the Fisher information. A trend of information tending to zero with diminishing of the details recently was aptly nicknamed "the Rayleigh catastrophe"  \cite{tsang2015quantum,lupo,nair2016far,tsang2016quantum,ang2017quantum}. However, also recently it has become obvious that the character of "the Rayleigh catastrophe" and the very fact of it occurrence are defined by the quantum states and correlations of the imaging field and measurements performed to register the image. Even for a set of incoherent point sources as an object, one can still devise measurements allowing to drastically improve precision in comparison with the common intensity distribution detection \cite{tsang2019quantum,zhou2019modern}, and even avoid "the Rayleigh catastrophe" for the case of estimating the distance between just two sources \cite{tsang2016quantum,ang2017quantum,paur2016achieving,rehacek2017optimal,vrehavcek2017multiparameter,vrehavcek2018optimal,tang2016fault}.

Resolution can be increased by using not only adequately designed measurements, but also classical and quantum correlations of the imaging field. A framework  to drastically improve resolution of an object by means of quantum image scanning microscopy is a combined measurement of intensity and intensity auto-correlations \cite{classen2017superresolution,tenne2019super}.
Illuminating the object with entangled quantum states started a novel subfield of research named "quantum radars" \cite{allen2008radar,lanzagorta2011quantum,jiang2013super,pirandola1}. Radar resolution gain, for example, was theoretically predicted \cite{pittman1995optical,tan2008quantum,guha2009gaussian} and experimentally  achieved \cite{lopaeva2013experimental,zhang2015entanglement} by "quantum illumination scheme", i.e.,  generating entangled states, retaining parts of these states, sending other parts to the object and implementing an entangling measurement of the scattered field with the retained one \cite{lloyd2008enhanced}, or implementing "ghost imaging" measurement, when the scattered part is just registered, but the retained part is scanned and both measurements results are correlated \cite{pittman1995optical,ferri2010differential,padgett2017introduction}. The use of optical principles to generate, control, and measure microwave and THz signals, widely known as microwave photonics \cite{marpaung2019integrated}, has been the area of intense research activities in recent years. Whereupon, quantum illumination strategy was also extended to the microwave region preferable for radar implementation due to lower losses and noise in comparison with the near-optical wavelengths \cite{barzanjeh2015microwave}.

The "ghost imaging" can be realized with classically correlated light beams albeit with smaller signal-to-noise ratio \cite{bennink2004quantum}. Methodologically, it resembles the classical scheme of so called "noise radars" \cite{guosui1999development}. A random noise signal is transmitted, and its delayed version is used as  reference. The signal scattered by the target is received and mixed with the reference. Then, the delayed correlation function of the reference and the signal exhibits a sharp maximum for the delay corresponding to the distance to the scattering object. Curiously, such a classical scheme can be enhanced by implementing a two-mode squeezed source as a signal and reference \cite{chang2019quantum}.

Here we consider the problem of increasing resolution of the object details by exploiting correlations of an imaging field. We suggest using both temporal and spatial correlations of the field scattered from the object for avoiding "the Rayleigh catastrophe" and enhancing resolution of the object configuration recovery. To achieve this one does not need sophisticated techniques such as, for example, measuring the image moments \cite{tsang2019quantum,zhou2019modern}. We demonstrate that for super-resolution, one can simply measure the delayed second-order correlation function of the scattered field in different spatial locations. For that one does not need a retained reference field, but the source field should possess specific spatiotemporal correlations. Here we show that even simplest quantum "talking" antenna consisting of two interacting two-level systems emitting into the open space is able to produce field with necessary correlations.  We demonstrate that for resolving two small scatterers in our scheme, "the Rayleigh catastrophe" is not present. For a larger number of scatterers, measuring the delayed second-order intensity correlation function allows to drastically enhance resolution in comparison with the measurement of the zero-delay intensity correlation function (the latter case being equivalent to measurement with field emitted by two uncorrelated dipoles). Moreover, our scheme is rather tolerant to uncertainty in detection time corresponding to a finite detection window of realistic detectors. Even uncertainty comparable to delay time does not spoil the super-resolution and does not return "the Rayleigh catastrophe".

We also consider an application of our scheme to near-field imaging and estimation of antenna parameters, such as a rotation angle and emitters separation. For these cases, one can also avoid "the Rayleigh catastrophe" for parameters in question tending to zero. However, for estimating the emitters separation time-uncertainty is much more destructive than for other considered cases, practically, restoring "the catastrophe".

\section{The measurement scheme and information}
Here we consider the  following simple sensing/imaging scheme. The quantum antenna source produces the field. This field  is transformed by object in the far-field zone and then propagates to the detectors able to measure field correlations between different spatiotemporal locations. We assume that results of our measurements  can be described by the set of probabilities $p_j({\vec x};{\vec y})$, where parameters ${\vec x}$ describe quantities to be inferred, and parameters ${\vec y}$ describe controlled parameters of the set-up. This set is assumed to be complete, $\sum\limits_{\forall j}p_j({\vec x};{\vec y})=1$. The information content of such a measurement can be conveniently described by the classical Fisher information matrix (FIM) with the following elements
\begin{eqnarray}
F_{kl}=\sum\limits_{\forall j}\frac{1}{p_j}\left(\frac{\partial}{\partial x_k}p_j({\vec x};{\vec y})\right)
\left(\frac{\partial}{\partial x_l}p_j({\vec x};{\vec y})\right).
\label{fisher}
\end{eqnarray}
If the estimate is unbiased, the variance $\Delta^2_k$ of $k$-th parameter   satisfies the Cramer-Rao bound (CRB)
\begin{equation}
\Delta^2_k\ge \frac{1}{N}[F^{-1}]_{kk}.
\label{cramer}
\end{equation}
where $N$ is the total number of registered outcomes. Thus, the total error described as the sum of variances (\ref{cramer}) is bounded as
\begin{equation}
\Delta^2=\sum\limits_{k=1}^M\Delta^2_k\ge \frac{1}{N}Tr[F^{-1}]\ge \frac{1}{Nf_{min}},
\label{cramertotal}
\end{equation}
where $M$ is a number of parameters to infer and $f_{min}$ is a minimal eigenvalue of the FIM (\ref{fisher}).

"The Rayleigh catastrophe" occurs, when for particular values of parameters $f_{min}=0$. So, near these particular values one needs using very large number of registered outcomes for achieving low inference errors. It is already well understood that "the Rayleigh catastrophe" is not a fundamental fact but rather a consequence of having  a particular type of measurements implemented for imaging \cite{tsang2015quantum,paur2016achieving}. For example, implementing  the centroid measurement instead of usual intensity measurement for inference of a distance between two incoherent point sources, it is possible to obtain a non-zero Fisher information for coinciding sources \cite{tsang2016quantum,ang2017quantum,paur2016achieving, rehacek2017optimal,vrehavcek2017multiparameter,vrehavcek2018optimal,tang2016fault}. For our arrangement avoiding "the Rayleigh catastrophe" means that if for some ${\vec x}_0$, ${\vec y}_0$ we have $f_{min}=0$, we should look for such ${\vec y}$ as to remove the FIM degeneracy for this ${\vec x}_0$. Generally, in our scheme looking for higher resolution means looking for the set of parameters ${\vec y}$ to decrease $\Delta^2$ for the same ${\vec x}$ and $N$.

\subsection{The quantum antenna source}
We consider the simplest "talking" quantum antenna of just two identical two-level systems (TLS) in a homogeneous vacuum. For simplicity sake, we assume that dipole moments are parallel to each other and orthogonal to the line connecting them. The  positive-frequency part of the field operator  that gives non-zero contribution to the normally ordered correlation functions and describes the spatial field distribution at the point ${\vec r}$ in the far field zone, reads as \cite{scully1999quantum}:
\begin{eqnarray}
\vec{E}({\vec r},\grave{t})\propto \sum\limits_{j=1}^2\frac{{\vec n}\times[{\vec n}\times {\vec d}]}{|{\vec r}|}
\exp\left\{-i\frac{\omega}{c}\frac{\vec r}{|{\vec r}|}{\vec R}_j\right\}\sigma^-_j(t),
\label{arop}
\end{eqnarray}
where $\sigma^-_j(t)$ is the time-dependent lowering operator for the $j$-th TLS with upper (lower) levels described by the vectors $|\pm_j\rangle$, ${\vec d}$ is the dipole moment vector; and ${\vec n}$ is the unit vector from an emitter to the observation point; the vector ${\vec R}_j$ describes the position of $j-$th TLS; $\omega$ is the TLS transition frequency, and $\grave{t}=t+|{\vec r}|/c$.

 To describe the field measurements results, i.e., averages of the field operators in the far-field zone, one needs finding correlation functions of the TLS operators for different time-moments. Using Markovian approximation, in the basis rotating with the frequency $\omega$, dynamics of the density matrix of these two emitters is described by the following  master equation \cite{scully1999quantum,breuer2002theory}:
\begin{align}
 \label{mast1}
 \nonumber
\frac{d}{dt}\rho=&-i{f_{12}}[\sigma_1^+\sigma_2^-+h.c.,\rho]+\\
&\frac{1}{2}\sum\limits_{j,l=1}^2\gamma_{jl}(2\sigma^-_j\rho\sigma^+_l-\sigma^+_j\sigma^-_l\rho-\rho\sigma^+_j\sigma^-_l),
\end{align}
where the unitary (elastic) coupling coefficient is
\begin{eqnarray}
f_{12}=-\frac{3}{2}\Gamma\Bigl(\frac{\cos\{\zeta_{12}\}}{\zeta_{12}}-
\Bigl(\frac{\sin\{\zeta_{12}\}}{\zeta_{12}^2}+\frac{\cos\{\zeta_{12}\}}{\zeta_{12}^3} \Bigr)\Bigr)
\label{f0}
\end{eqnarray}
and the dissipative (inelastic) coupling coefficients are
\begin{gather}
\label{gam1}
\gamma_{11}=\gamma_{22}=\Gamma,  \\
\nonumber
\gamma_{12}=\frac{3}{2}\Gamma\left(\frac{\sin\{\zeta_{12}\}}{\zeta_{12}}-
\Bigl(\frac{\sin\{\zeta_{12}\}}{\zeta_{12}^3}-\frac{\cos\{\zeta_{12}\}}{\zeta_{12}^2} \Bigr)\right),
\end{gather}
where $\Gamma={\omega^3|\vec d|^2}/{3\hbar\pi\varepsilon_0c^3}$ is the spontaneous emission rate for the single TLS; $\zeta_{12}=\omega \zeta/c$ with $\zeta$ being the distance between the emitters.

Notice that  in difference with the standard Dicke model, Eq.(\ref{mast1}) predicts appearance of both the super-radiant and sub-radiant components of the population decay of both TLS even in the case when both TLS are initially completely excited \cite{ficek1987quantum,ficek1988quantum}.
Moreover, in difference with the standard Dicke model predicting no entanglement between TLS during the superradiant decay  of the completely excited initial state \cite{wolfe2014certifying}, the entanglement does appear between TLS separated with the distance comparable with the resonant wavelength \cite{tanas2004entangling}.

\subsection{The object and the measurement}

For simplicity sake, here we consider field propagation in the plane perpendicular to dipoles, and assume that detectors are also in this plane. So, we can describe the field (\ref{arop}) just by the scalar amplitude, and assume that the field transformation toward the observation point is described by the following linear transformation
\begin{eqnarray}
{E}_{out}({\vec r},\grave{t})=\sum\limits_{j=1,2}{f_j}({\vec r},{\vec x})\sigma^-_j(t),
\label{obj}
\end{eqnarray}
where the operator $\vec{E}_{out}({\vec r},t)$ describes the field at the observation point ${\vec r}$. The functions  ${\vec f_j}({\vec r},{\vec x})$ are defined by the imaged objects and the measurement set-up. In the subsequent Sections we consider several examples of the measurement schemes and corresponding functions ${\vec f_j}({\vec r},{\vec x})$.

For capturing temporal and spatial correlations of the emitted photons pairs, we suggest measuring the delayed second-order intensity correlation function
\begin{align}
\nonumber
G^{(2)}(\theta_1,\grave{t};\theta_2,\grave{t}+\tau)=& \sum\limits_{j,l,m,n=1}^2 F_{j,l,m,n}(\theta_1,\theta_2)\times \\
&{\langle}\sigma_j^+ (t)\sigma_l^+ (t+\tau)\sigma_m^- (t+\tau)\sigma_n^- (t){\rangle},
\label{g2}
\end{align}
where $\theta_j$ are observation angles, the function $F_{j,l,m,n}(\theta_1,\theta_2)$ is defined from Eq.(\ref{obj}) and $\tau\geq0$ is the delay between counts on pair of the detectors.
Four-operator correlation functions of Eq.(\ref{g2}) can be written in the following way
\begin{multline}
{\langle}\sigma_j^+ (t)\sigma_l^+ (t+\tau)\sigma_m^- (t+\tau)\sigma_n^- (t){\rangle} =\\
\frac{1}{2} \exp\left\{-\Gamma(2t+\tau)\right\}\Upsilon_{jn}^{lm}(\tau),
\label{4s}
\end{multline}
where
\begin{eqnarray}
\Upsilon_{jn}^{jn}(\tau)
= -\cos\left\{f_{12}\tau\right\}+\cosh\left\{\gamma_{12}\tau\right\},
\label{up1}
\end{eqnarray}
for $j,n=1,2$. For $j\neq n$ we have
\begin{align}
\nonumber
\Upsilon_{jn}^{nj}(\tau)&=\Upsilon_{jj}^{nn}(\tau)=
\cos\left\{f_{12}\tau\right\}+\cosh\left\{\gamma_{12}\tau\right\}, \\
\label{up2}
\Upsilon_{jn}^{jj}(\tau)&=\Upsilon_{jj}^{jn}(\tau)=-i\sin\left\{f_{12}\tau\right\}-\sinh\left\{\gamma_{12}\tau\right\},\\
\nonumber
\Upsilon_{jj}^{nj}(\tau)&=\Upsilon_{nj}^{jj}(\tau)=i\sin\left\{f_{12}\tau\right\}-\sinh\left\{\gamma_{12}\tau\right\}.
\end{align}

Notice that the correlation function
(\ref{g2}) is proportional to the photon count rate per unit solid angle of having the first photon  registered at the observation angle $\theta_1$ at the time-moment $\grave{t}$ and the second photon registered at the observation angle $\theta_2$ at the time-moment $\grave{t}+\tau$.

\begin{figure}[htb]
\includegraphics[width=\linewidth]{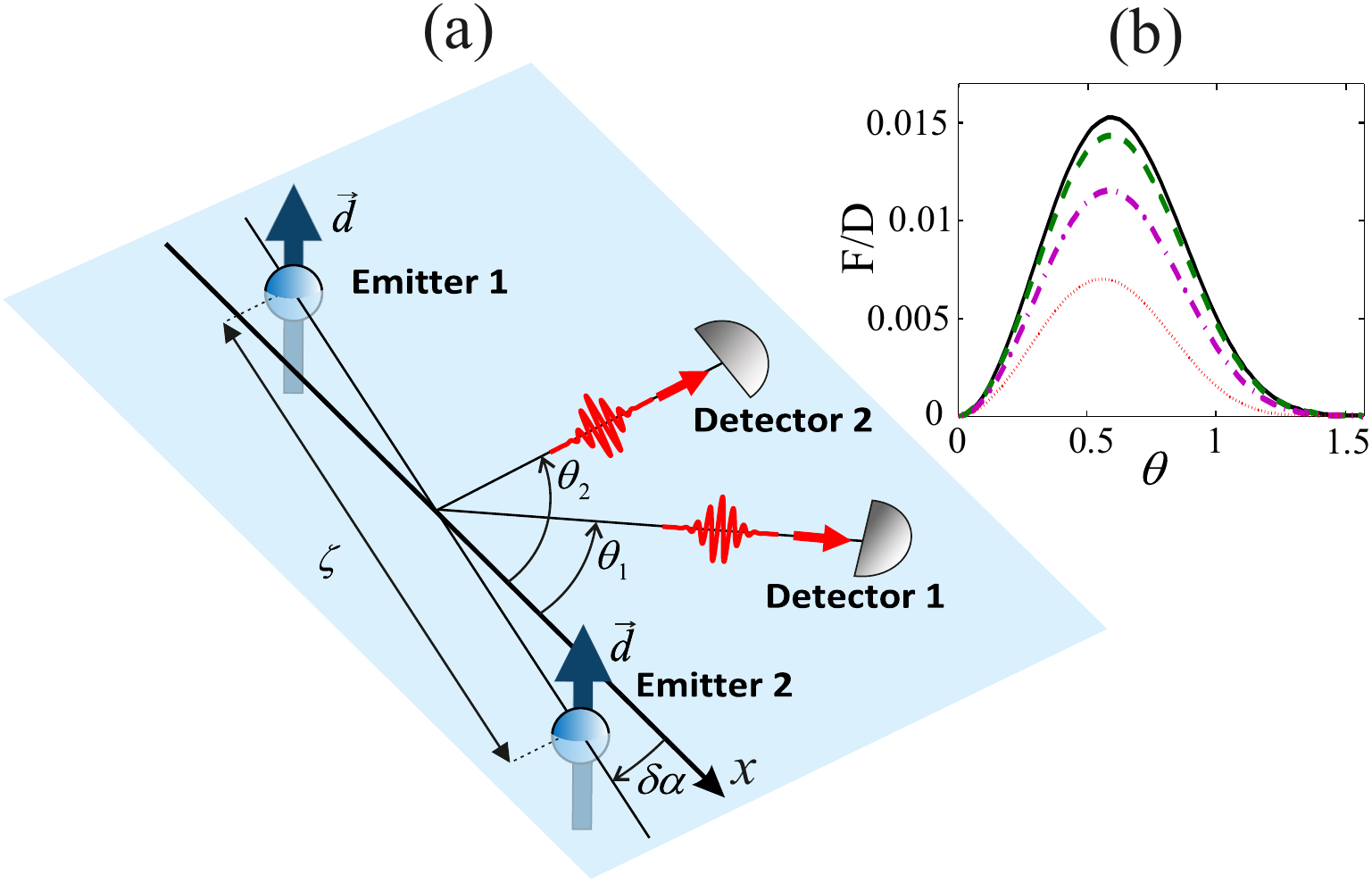}
\caption{(a) The scheme of the simplest quantum antenna for producing correlated photon pairs registered in the far-field zone on two detectors seen at angles $\theta_1$ and $\theta_2$ from the $x$ axis. The angle $\delta\alpha$ denotes antenna rotation. (b) The  Fisher information for detection of a zero rotation angle $\delta\alpha$ in dependence on the direction angle $\theta$ as given by Eq.(\ref{prot2})
for the delay $\Gamma\tau=0.75$. Solid, dashed, dash-dotted and dotted lines correspond to the following averaging times $\Gamma\delta\tau=0,0.25,0.5,0.75$. For this picture the distance between dipoles $\lambda/2\pi=1$.} \label{fig1}
\end{figure}

\section{The antenna rotation detection}

First of all, let us consider a specific simple far-field measurement scheme which will serve us as a basis for subsequent considerations illustrating the role of the delay. This scheme is just a detection of the second-order intensity correlation function $G^{(2)}(\theta_1,t;\theta_2,t+\tau)$ for sensing the rotation of the antenna. The scheme of the measurement is shown in Fig.\ref{fig1}: two point detectors are in the plane perpendicular to the dipoles; observation angles $\theta_{1,2}$ are measured from the $x$ axis.  We are looking for a small rotation of the antenna axis $\delta\alpha$ in the plane orthogonal to the dipoles direction. For the case Eq.(\ref{obj}) can be re-written as
${E}(\theta_i,\grave{t})=f_1\sigma_1^-(t)+f_2\sigma_2^-(t)$,
where
$f_1\propto const$, $f_2\propto \exp\left\{-i\zeta_{12}\cos \left\{\theta_i+\delta \alpha \right\}\right\}$.

To demonstrate the importance of the delay for the measurement, we take the same observation angles, $\theta_1=\theta_2=\theta$.
Then, the probability to detect the first photon in a small time window near the moment $t$ on the first detector and the second photon near the moment $t+\tau$ on the second detector can be written as
\begin{align}
\nonumber
p(\theta;\tau)=D\exp\{-\Gamma\tau\}\times \rule{2.2cm}{0pt}
&\\
\nonumber
\label{prot1}
\Biggl[\Bigl(1+\cos^2\phi  (\theta+\delta\alpha)\Bigr)\cosh\{\gamma_{12}&\tau\}-\\
\nonumber
2\cos\phi  (\theta+\delta\alpha)\sinh\{\gamma_{12}&\tau\}+\\
\sin^2\phi  (\theta+\delta\alpha)\cos\{f_{12}&\tau\} \Biggr],
\end{align}
where the multiplier $D$ depends on the area of detectors  and their efficiencies, the size of the time-window for the detection, the distance from the antenna and the time-moment of the registration; the angle $\phi(\theta)=\zeta_{12}\cos\theta$.

Eq.(\ref{prot1}) clearly shows that  only presence of a finite delay $\tau>0$ enables one to detect a rotation by a single-direction measurement. For the zero delay the probability (\ref{prot1}) loses spatial dependence appearing for different observation angles, $\theta_1\neq\theta_2$ (in that case $G^{(2)}(\theta_1,\grave{t};\theta_2,\grave{t})\propto 1+\cos{\{\phi(\theta_1)-\phi(\theta_2)\}}$  ).  Also, presence of the delay allows one to detect arbitrary small antenna rotation without having an error tending to infinity.

Let us demonstrate it for just a single measurement for coinciding observation angles. The detectors are small and distant from the antenna, so, one can safely assume that $p(\theta;\tau)\ll 1$. Assuming that we are making just one measurement, our set of probabilities would be $p$ and $1-p\gg p$.  We take that the detectors have an uncertainty in opening the time-window. So, after the first click at some moment $t$, the time-window for the second detector is opened not exactly at the chosen moment $t+\tau$, but in some interval around it. Thus, the corresponding Fisher information  reads as
\begin{align}
\nonumber
F&\approx \frac{D}{{\bar p}(\theta;\tau)}\left(\frac{\partial}{\partial\theta}{\bar p}(\theta;\tau)\right)^2, \\
{\bar p}(\theta;\tau)&=\int\limits_{\tau-\delta\tau}^{\tau+\delta\tau}dx \Omega(x) p(\theta;x),
\label{prot2}
\end{align}
where the distribution $\Omega(x)$ describes the probability of different delays.
Fig.\ref{fig1} (b) shows an example of the Fisher information corresponding to $\delta\alpha=0$ for the delay $\Gamma\tau=0.75$ and different averaging intervals (starting from the sharp measurement) as given by Eq.(\ref{prot2}) for $\Omega(x)$ being homogeneous in the time-interval $[\tau_0-\delta\tau,\tau_0+\delta\tau]$ around $\tau_0>0$. The Fisher information is expectedly equal to zero at the angles $\theta=0$ and $\pi/2$ (since the differentiation of $p(\theta;\tau)$ give a result proportional to $\sin\theta$) but is not equal to zero in between. Moreover, even large uncertainty of the second photon detection does not spoil the effect. The Fisher information diminishes but remains finite.

\begin{figure}[htb]
\includegraphics[width=\linewidth]{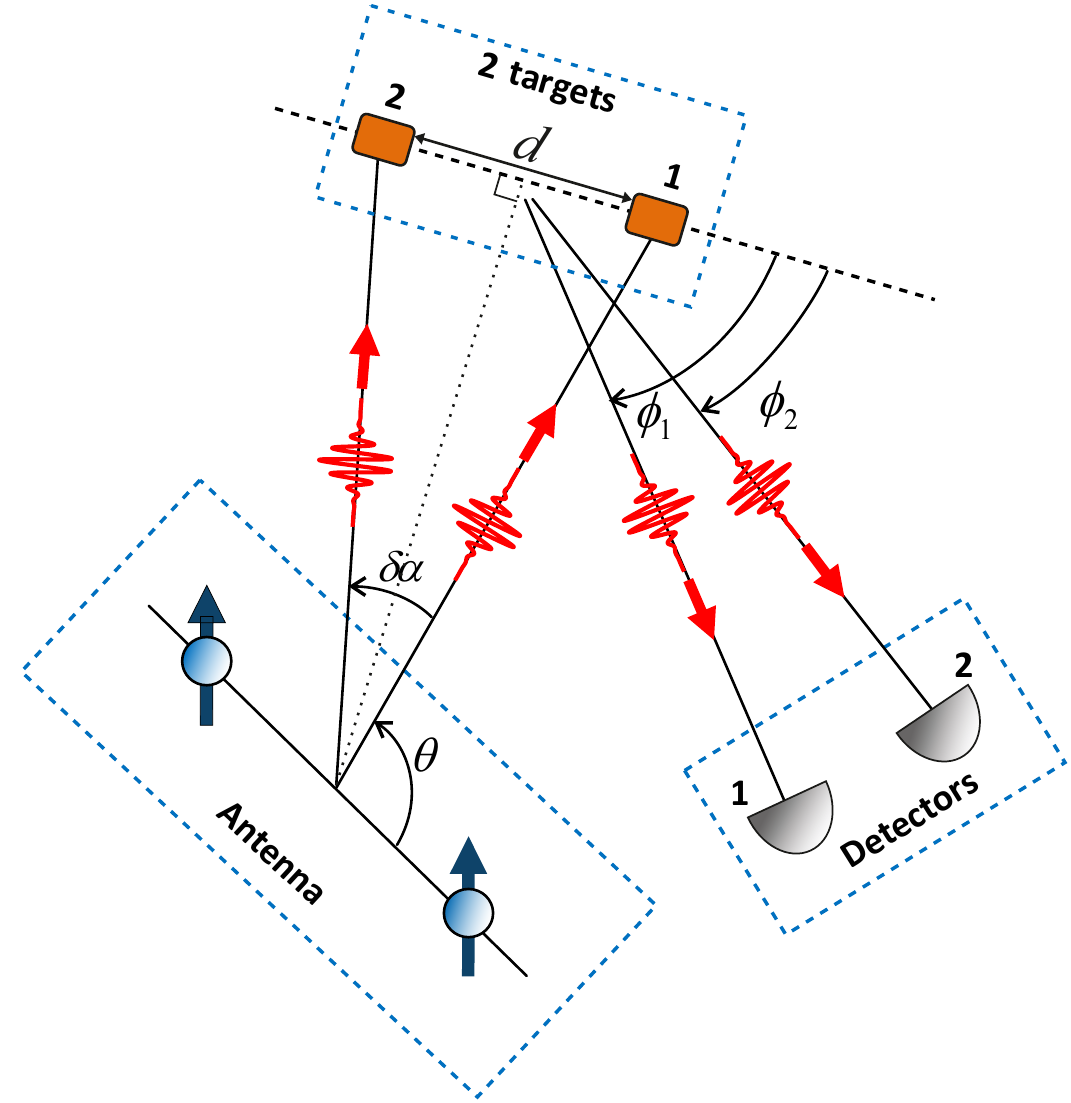}
\caption{The scheme of the  far-field scattering imaging with the simplest quantum antenna. The field produced by the antenna in the far-field zone impinges on set of targets, and the scattered field is detected in the far-field zone by two detectors, ${\mathrm D_1}$ and ${\mathrm D_2}$.}
\label{fig3}
\end{figure}

\begin{figure}[htb]
\includegraphics[width=\linewidth]{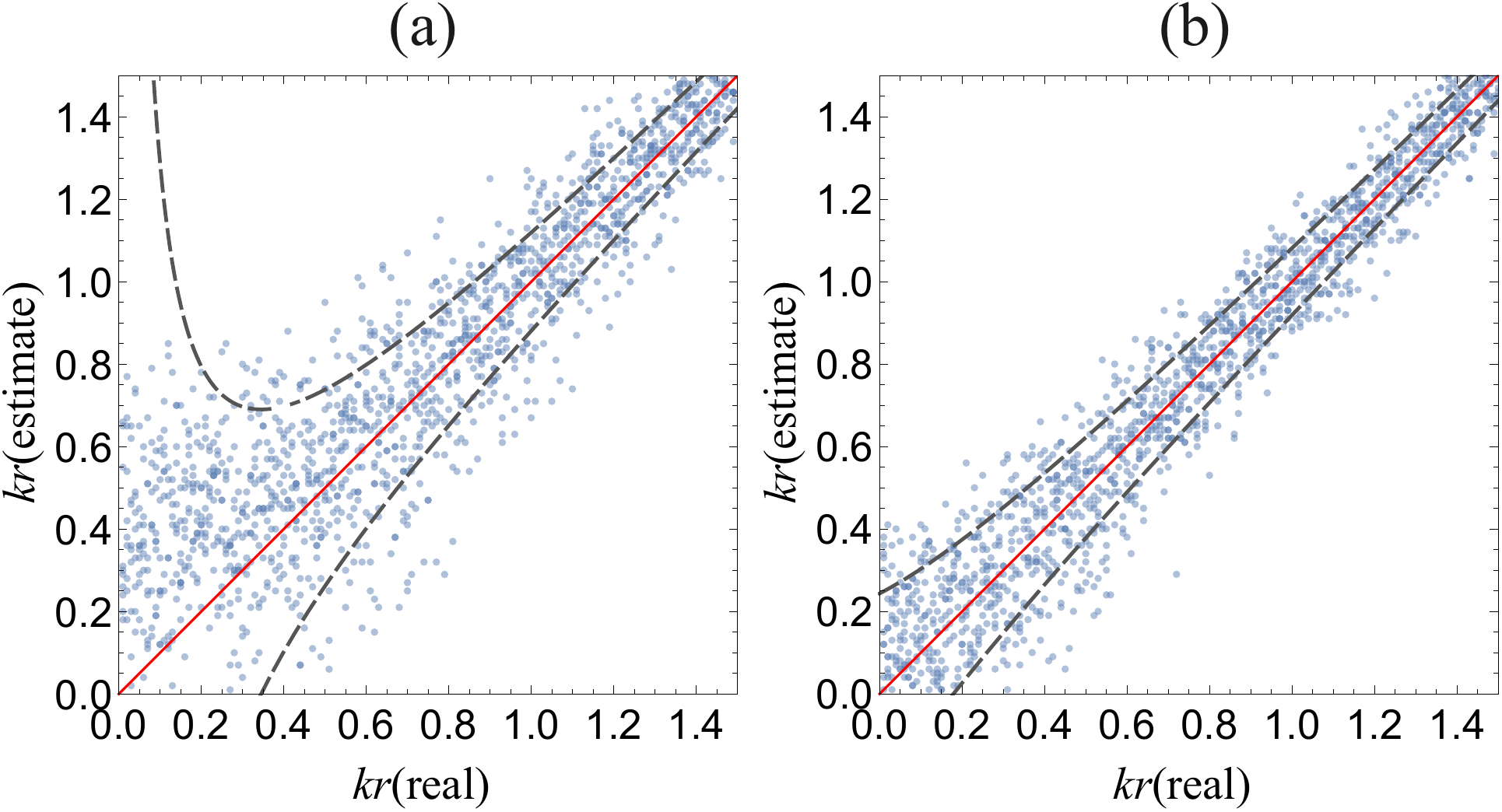}
\caption{An illustration of the reconstruction of the distance between scatterers for the zero-delay measurements (a) and for the delayed $G^{(2)}$ measurements in dependence on this distance. Dashed lines depict the lower bound on errors as predicted by the Fisher information for $N=10^7$ counts per measurement. The red line is the guide for the eye showing true separation between objects. For both panels $\zeta_{12}=4.8$, the angle between axes of the antenna and direction to the object is 1.2 rad, observation angles for the first and second measurements are $\pi/2.15$ and  $\pi/2.15 + \pi/50$. For the panel (b) delays for the first and second measurements are  $\Gamma \tau=0.5,0.75$ respectively. The Fisher information was calculated for the probabilities normalized by their sum.}
\label{fig4}
\end{figure}

\section{The far-field scattering imaging}

 Now let us consider the radar-like scheme exploiting noise intensity correlations of the scattered field. We discuss far-field sensing with the field generated by our quantum antenna. We  demonstrate how "the Rayleigh catastrophe" arises and how the measurement of  the delayed $G^{(2)}$ removes it. We also demonstrate that registering the delayed intensity correlations gives a possibility to drastically increase resolution in comparison with the zero-delay detection.

\subsection{A simple model: resolving two point scatterers}

Here we consider an example of just two scatterers and demonstrate how "the catastrophe" arises and how one can remove it. Our radar-like set-up is shown in Fig.\ref{fig3}. The field emitted by our antenna is scattered by two point-like  objects toward two detectors in the far-field zone of the scattered field, and there it is registered by two detectors for inferring the delayed second-order correlation function. Both the objects and the detectors are in the plane perpendicular to dipole moments. Object axis is orthogonal to the illumination direction (the line connecting the antenna with the object). We aim to infer the distance between the scatterers by measuring the delayed second-order correlation function of the scattered field. The positive-frequency part of the scattered field amplitude in the far-field zone at the position of $l$-th detector  can be represented as
\begin{multline}
\label{far1}
{E}(\phi_l,\acute{t})\propto \frac{\exp\{ikR_s\}}{R_s}\times\\
\left({E}({\vec r_1},\grave{t})+{E}({\vec r_2},\grave{t})\exp\{-idk\cos\phi_l\}\right)=\\
f_1(\phi_l)\sigma_1^-(t)+f_2(\phi_l)\sigma_2^-(t),
\end{multline}
where $R_s$ is the distance from the scatterer to the observation point (the detector); $\phi_l$ is the observation angle toward detector with respect to the line connecting the scatterers, and ${E}({\vec r_m},\grave{t})$ is the impinging field amplitude at the position of $m$-th scatterer; $\acute{t}=\grave{t}-R_s/c\geq0$; $d\approx R_o\delta\alpha$ is the distance between scatterers to be inferred; $R_o$ is the distance from the antenna to the scattering object. We take the object to be small, $\delta\alpha\ll1$. Assuming that the positions of the detectors are coinciding, we have the following expressions for the values of functions $f_n$
\begin{align}
\label{f2}
\nonumber
f_1(\phi_l)\propto \, & 1+\exp\{-ikR_o\delta\alpha\cos\phi_l\}, \\
f_2(\phi_l)\propto \, & \exp\{-i\zeta_{12}\cos\theta\} +\\
\nonumber
&\exp\{-i\zeta_{12}\cos\{\theta+\delta\alpha\}\}\exp\{-ikR_o\delta\alpha\cos\phi_l\}.
\end{align}

The considered far-field imaging scheme can be reduced to just the measurement of the second-order intensity correlation function in the far-field as it is for the previous Section. Indeed, for the observation direction orthogonal to the line connecting the scatterers, $\cos\phi=0$. Let us assume that the object is small, $\delta\alpha\rightarrow 0$. Thus, we shall have   $f_2(\pi/2)\propto 2\exp\{-i\zeta_{12}\cos\{\theta+\delta\alpha/2\}\}$.
The two-photon registration probability for the case coincides with the probability (\ref{prot1}) for the rotation detection for the rotation angle $\delta\alpha/2$. Just like for the rotation detection, there is no spatial dependence for zero delay. For the finite delay and for the single measurement as it is described in the previous Section, the Fisher information for $\delta\alpha\rightarrow0$ coincides with the one obtained for the rotation angle detection up to the constant multiplier. So, "the Rayleigh catastrophe" is absent, and the scheme is robust with respect to the delay uncertainty.

Notice, however, that in difference with the rotation detection, for zero delay "the Rayleigh catastrophe" is present for arbitrary observation angles. For zero delay Eqs.(\ref{g2})--(\ref{up2}) and functions  (\ref{f2}) lead to the following result for the observation angles $\phi_1$, $\phi_2$:
\begin{align}
p(\theta;0)=D\left|f_1(\phi_1)f_2(\phi_2)+f_1(\phi_2)f_2(\phi_1)\right|^2.
\label{pfar}
\end{align}
For $\delta\alpha\rightarrow 0$, Eq.(\ref{pfar}) shows that  $\frac{d}{d\delta\alpha}p(\delta\alpha,0)\rightarrow0$, and the Fisher information tends to zero as $\delta\alpha^2$, i.e., just like in the archetypal  case of resolving two-point sources by the image of the intensity distribution in the far-field \cite{tsang2015quantum,nair2016far, tsang2016quantum,ang2017quantum}.

To see how the minimal error bound is connected with the practical estimation of the distance between scatterers, we simulated reconstruction of the distance from modeled signal, represented by  frequencies of two measurements with different observation angles, both for the zero-delay (Fig.\ref{fig4}(a)) and a finite delay (Fig.\ref{fig4}(b)). Each estimation point corresponds to different distance between the dipoles, $kr$, scanned with a small step. Blue lines denote the bounds given by the Cramer-Rao inequality (\ref{cramertotal}). The number of measurement runs is the same for both panels. One can see that
the estimation performed by minimizing the distance between the reconstructed probability and the simulated relative frequency conforms well to the bound (\ref{cramertotal}). "The Rayleigh catastrophe" is manifested by respective increase of errors for smaller distances, whereas for the delayed measurements errors remain more or less the same.

\begin{figure}[htb]
\centering
\includegraphics[width=.9 \linewidth]{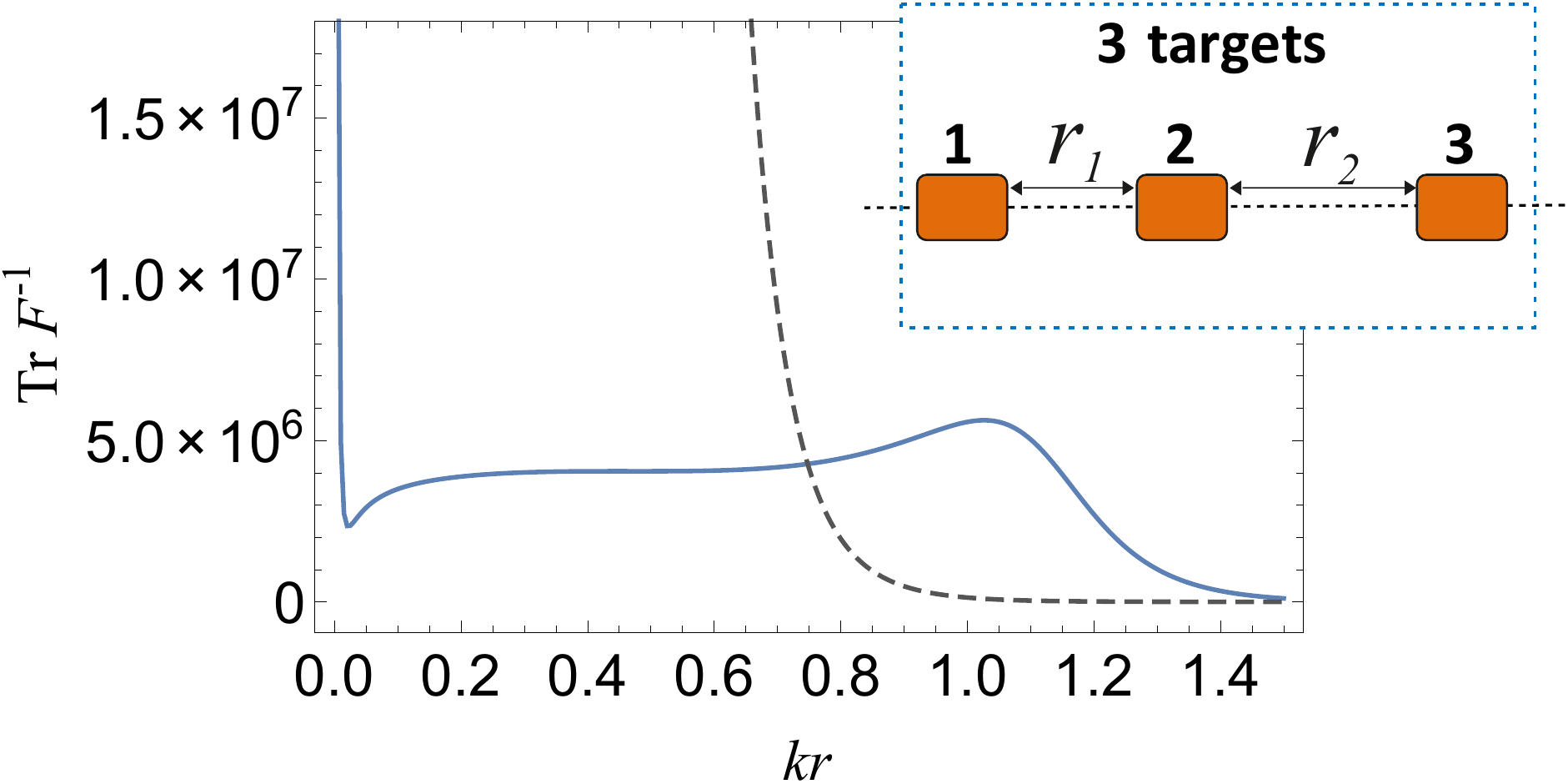}
\caption{Inset shows the three-scatterers system, $r_1$ and $r_2$ are distances to be inferred. The total error bound defined by the trace of the inverse Fisher information matrix corresponding to $r_1=r_2/2=r$ is shown for different values of the parameter $kr$, for the zero-delay measurements (dashed line) and the delayed measurements (solid line).  For both panels $\zeta_{12}=4.8$; the angle between axis of the antenna and the object is 1.2 rad; two detectors for the same observation angle were considered for $\phi_1=\phi_2=\pi/4,\pi/3,\pi/2,3\pi/4$. For the delayed measurements the following delays were taken $\Gamma\tau=1.5,2.5,2,0$.
The Fisher information was calculated for the probabilities normalized by their sum. }
\label{fig5}
\end{figure}

\subsection{Resolving several scatterers}

As it was already mentioned here, "the Rayleigh catastrophe" can be removed for the case of resolving just two small objects and just one inferred parameter, but for more complicated object structure and more parameters to infer, it returns \cite{tsang2019quantum,zhou2019modern}. However, measuring the delayed correlation function is able to bring about a considerable improvement of the resolution in comparison with the zero-delayed measurement. This improvement can have quite a drastic character. To demonstrate it, let us consider imaging of objects composed of just three small scatterers (Fig.\ref{fig5}(a)). The field impinging on the detector at the direction angle $\phi_l$ can be written as
\begin{equation}
\label{far3}
{E}(\phi_l,\acute{t})\propto \frac{\exp\{ikR_s\}}{R_s}
\sum\limits_{j=1}^3{E}({\vec r_j},\grave{t})\exp\{-id_{j-1}k\cos\phi_l\},
\end{equation}
where $d_{1}=r_1$, $d_2=r_1+r_2$; and $r_{1,2}$ are distances between the first and the second scatterers, and between the second and the third scatterers correspondingly; $d_0\equiv0$. The vectors ${\vec r_j}$ describe positions of the scatterers.

Examples of the total error bounds  (\ref{cramertotal}) for the delayed $G^{(2)}$ measurements and for zero-delay measurements are shown in Fig.(\ref{fig5}). Four measurement settings were considered for both kinds of measurements. Two detectors for the same angle were considered for
$\phi_1=\phi_2=\pi/4$, $\pi/3$, $\pi/2$, $3\pi/4$. For the delayed measurements the following delays were taken: $\Gamma\tau=1.5$ (for $\phi_1=\phi_2=\pi/4$), 2.5 ($\pi/3$), 2 ($\pi/2$), 0 ($\pi/4$).
For calculating the Fisher information matrix the normalized probabilities were introduced as $p_j/\sum\limits_{\forall m}p_m$; the variables to be inferred are $r_{1,2}$. The bound was estimated for $r_1=r_2/2$. One can see that the regions of the object size values where the total error  bound starts to strongly increase (and these values correspond to the classical Abbe limit \cite{helstrom1970resolvability}) are quite different for the delayed and zero-delay measurements.  The delayed measurement leads to the order of magnitude resolution increase.

\subsection{Influence of the antenna size}

\begin{figure}[htb]
\includegraphics[width=\linewidth]{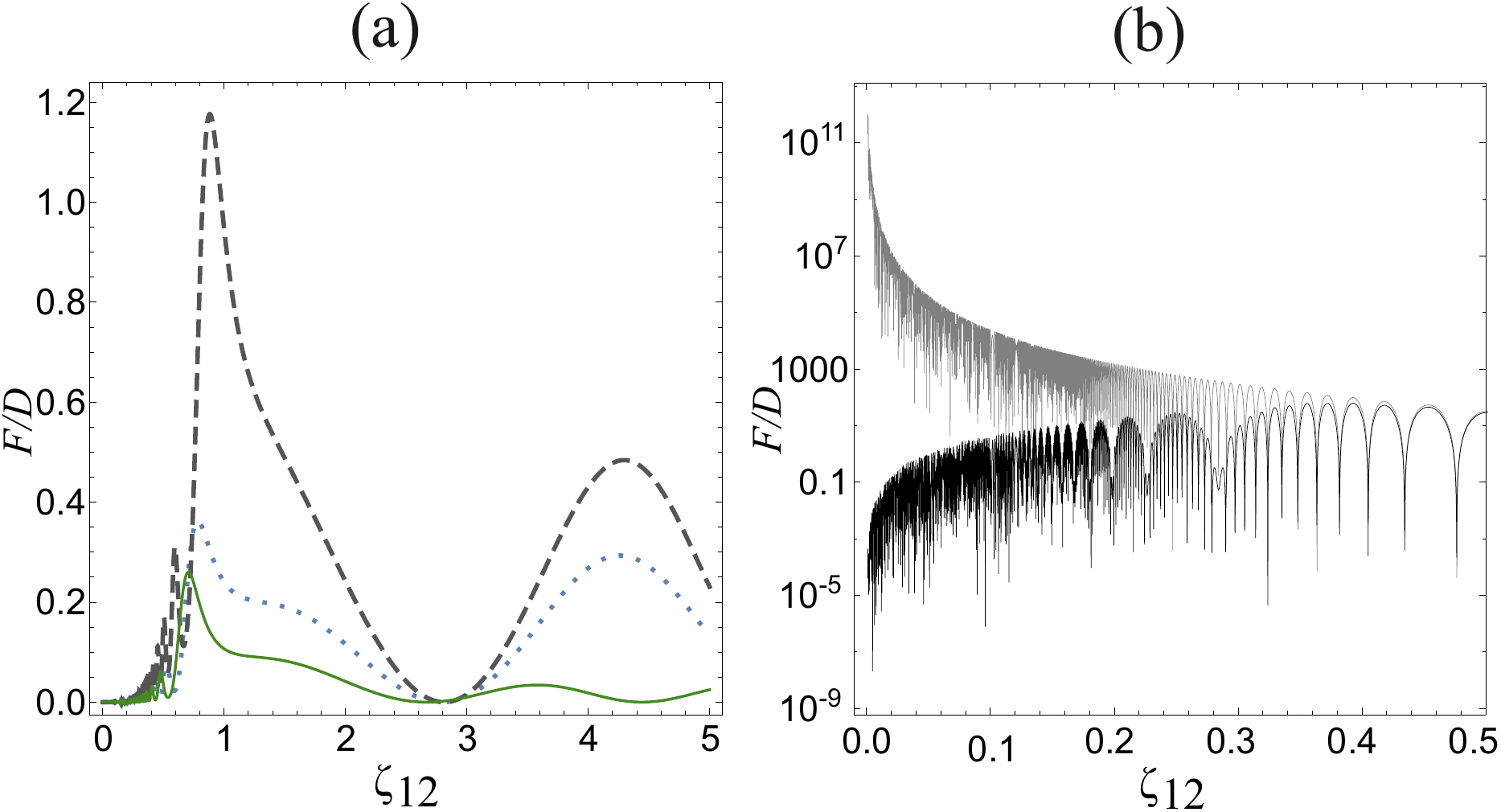}
\caption{(a) The Fisher information  for detection the distance between scatterers with a single-angle measurement for different delays and angles $\theta$ between axes of the antenna and the object.  in dependence on the distance between TLS. Solid, dotted, and dashed lines correspond to the following delays and observation angles $\Gamma\tau=0.75,\theta=\pi/4$, $\Gamma\tau=1.0,\theta=\pi/3$, $\Gamma\tau=1.5,\theta=\pi/3$. (b) The Fisher information for scattered far-field detection in dependence on the inter-dipole distance, $\zeta_{12}$. Gray and black lines correspond to the averaging time $\Gamma\delta\tau=0$ and 0.05. For all panels the delay $\Gamma\tau=0.75$.}
\label{fig6}
\end{figure}

The information provided by our correlation detection scheme depends on the distance between the antenna TLS rather nontrivially.  Eqs.(\ref{f0},\ref{gam1}) show that interaction between TLS intensifies when the distance between TLS becomes smaller. However, it can lead to worsening of the resolution, and to eventual return of the "the Rayleigh catastrophe" for the delayed measurements. This situation is illustrated in
Fig.\ref{fig6}. For different parameters of the scheme (delays, angles between the antenna axis and the line connecting the scatterers) the Fisher information invariably tends to zero for $\delta\alpha\rightarrow0$ with diminishing of the distance $\zeta_{12}$ between antenna TLSc proportionally to $\zeta_{12}^2$. However, the Fisher information can remain rather large even for such distances between antenna TLS that the interaction between them is already quite small, $|\gamma_{12}|,|f_{12}|\ll\Gamma$ (Fig.\ref{fig6}). Advantages of large phase shift provided by the larger $\zeta_{12}$ partially compensate the effect of decreasing $|\gamma_{12}|,|f_{12}|$.

Here we do not consider the case of different TLS of the antenna. However, it is not hard deriving the Fisher information from the analytic solution for two different TLS given in Res.\cite{ficek1987quantum,ficek1988quantum} for our detection scheme. While it is not possible avoiding "the catastrophe" for  $\zeta_{12}\rightarrow0$, it still possible for a fixed small $\zeta_{12}\ll1$ to increase the Fisher information by orders of magnitude by taking different decay rates $\Gamma$s for different TLS. This opens possibilities of using such small antenna-like arrangements, as, for example, color centers in diamond coupled to nano-cavities \cite{evans2018photon}.

\begin{figure}[htb]
\includegraphics[width=\linewidth]{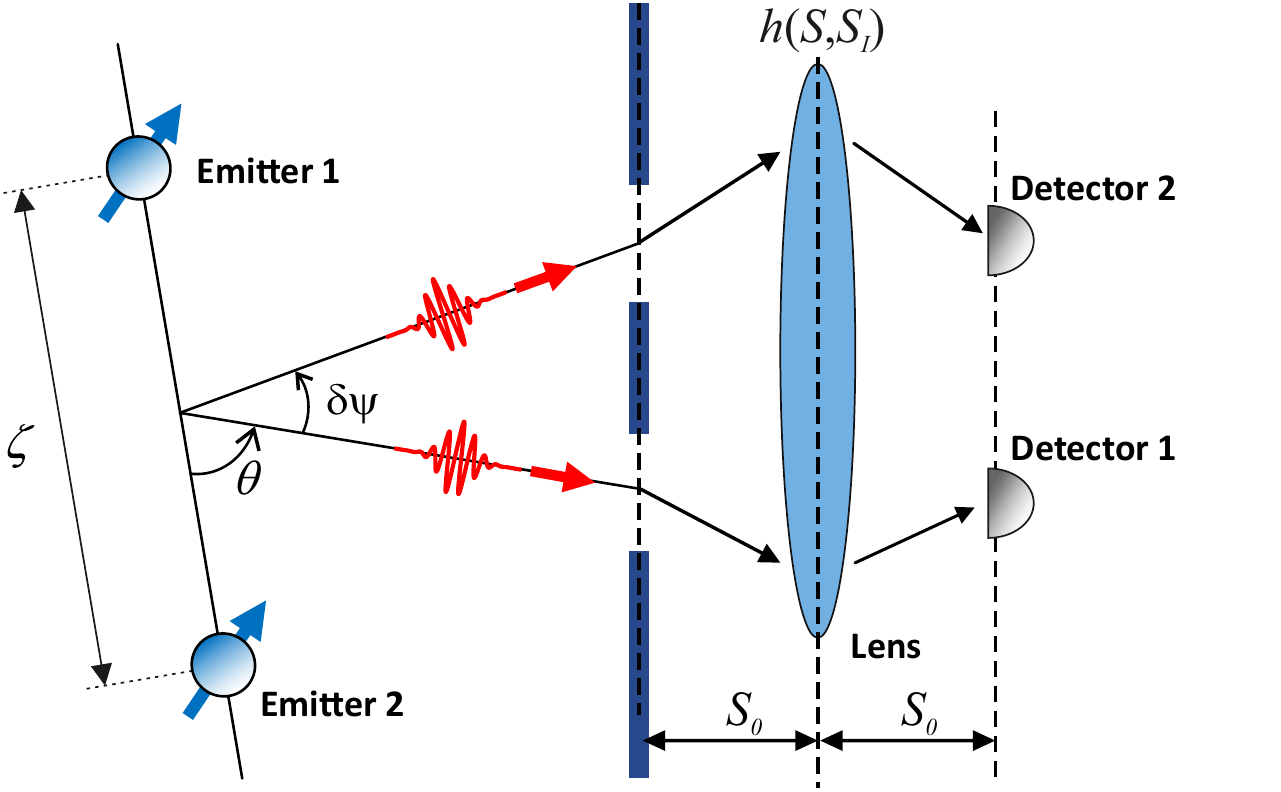}
\caption{The scheme of the near-field imaging with the simplest quantum antenna. The field produced by the antenna of the size $\zeta$ in the far-field zone impinges on the two small holes in the object plane and propagates toward the image plane. The direction angles toward these holes are $\theta+\delta\psi$ and $\theta$. This plane is in the near-field zone with respect to the object plane, the function $h({\vec S},{\vec S}_I)$ describes the field propagation between planes. Then, the  field is detected by two detectors, $S_0$ are distances between the lens and the image plane, and between the object plane and the lens.}
\label{fig7}
\end{figure}

\section{Other imaging schemes}

Correlations produced by the quantum antenna and exploited by the measurement of delayed $G^{(2)}$ can be used for other schemes of sensing and imaging. Here we analyze possibilities of using the antenna for super-resolving near-field imaging and estimating the distance between TLS, $\zeta_{12}$, by registering $G^{(2)}$ in the far-field zone.

\subsection{Near-field imaging}

 We consider a standard object used for illustration of the resolution: just two spatially separated small holes with the distance $r$ between them (see Fig.\ref{fig7}). After impinging on our object, the field propagates toward the image plane, and the positive-frequency part of its amplitude is given by the operator
\begin{multline}
{E}_I({\vec S}_I,t)=\int\limits_{O}d^2S h({\vec S}_I,{\vec S}){E}({\vec S},t)\approx \\
o \left(h({\vec S}_I,{\vec S}_1){E}({\vec S}_1,t)+ h({\vec S}_I,{\vec S}_2){E}({\vec S}_2,t)\right)=\\
f_1({\vec S}_I)\sigma_1^-(t)+f_2({\vec S}_I)\sigma_2^-(t),
\label{nearimag1}
\end{multline}
where the function $h({\vec S}_I,{\vec S})$ describes the field propagation between planes, and $o$ is the area of each hole. The field passing through holes is collected by a lens and is registered by the point detectors \cite{d2005quantum}. We assume that both the detectors are at the point ${\vec S}_I=0$, which is located directly opposite  the hole seen at the direction angle $\theta$ from the antenna. Thus, for the values of functions $f_j({\vec S}_I)$  we have the following expressions
\begin{align}
\label{f1}
f_1\propto \, & 1+\mathrm{somb}\left(\delta\psi x\right), \\
\nonumber
f_2\propto \,  & \exp\{-i\zeta_{12}\cos\theta\}+ \\
\nonumber
& \exp\{-i\zeta_{12}\cos\{\theta+\delta\psi\}\}\mathrm{somb}\left(\delta\psi x\right),
\end{align}
where $\delta\psi$ is the angle between directions toward the object holes; $r\approx \delta\psi R_o$, where $R_o$ is the distance from the antenna to the both holes. The function $\mathrm{somb}(y)=2J_1(y)/y$, $J_1(y)$ is the Bessel function of the first kind. The parameter $x=L\omega R_o/cs_0$, where $L$ is the lens radius and $s_0$ is the distance from the object plane to lens, and from the lens to the image plane.

Eqs.(\ref{f1}) point to the fact that near $\delta\psi=0$ the Fisher information is quite similar to the one obtained in the previous Section for the far-field sensing (and for detection of the antenna rotation). Indeed, $\mathrm{somb}(y)=1$, and $\frac{d}{dy}\mathrm{somb}(y)=0$ for $y=0$. Setting the sombrero functions in Eqs.(\ref{f1}) to unity gives one the values $f_j$ the same as they are for detection of $G^{(2)}(\theta,t;\theta+\delta\psi,t+\tau)$  without the imaging scheme. So, for the considered detection scheme the Fisher information for the coinciding holes, $r=0$, behaves just the like the one obtained in the previous Section.

\subsection{Inferring parameters of the antenna}

Finally, let us consider the task of inferring the parameters of the antenna by measuring the delayed $G^{(2)}$. We seek to find the distance, $\zeta$, between the dipoles. One can surmise that the behaviour of the Fisher information is hardly trivial for the case, since the rate of the unitary excitation exchange between emitters tends to infinity with the distance tending to zero, $|f_{12}(\zeta_{12})|\rightarrow\infty$ for $\zeta_{12}\rightarrow0$ (see Eq.(\ref{f0})). Also, $\frac{\partial}{\partial\zeta_{12}}\cos(f_{12}\tau)\rightarrow\infty$ for $\zeta_{12}\rightarrow0$ and $\tau\neq0$. So, the derivative of the probability (\ref{prot1}) diverges for $\zeta_{12}\rightarrow0$, and the Fisher information diverges, too. However, it is easy to see that in reality such an information divergence is not taking place. First of all, approximations leading to Eq.(\ref{f0})  eventually breaks down for too small distance between the emitters (even the rotating-wave approximations are not working for $|f_{12}(\zeta_{12})|\sim\omega$). Also, even an arbitrary small but finite averaging over the second photon registration moment removes the divergence. This is illustrated in Fig.\ref{fig6}(b) for the ideal time-sharp detection, and for the time-averaged case. One can see that even for quite small time-averaging ($\Gamma\delta\tau=0.05$ for the black line in Fig.\ref{fig6}(b)) the FIM behaves quite differently from the time-sharp case. Reality restores the Rayleigh catastrophe.

\section{Conclusions}

We have demonstrated that the simplest quantum "talking" antenna of just two interacting two-level systems is able to produce a field state useful for achieving super-resolution in far-field sensing. We have analyzed a quantum version of the noise radar scheme. The intensity correlation function of  the field scattered by the object is registered for different delay times. We have shown that one can achieve a super-resolving estimation of the scattering objects details in comparison, for example, with sensing using uncorrelated antenna sources. The delayed measurement allows to avoid "the Rayleigh catastrophe" for resolving two small scatterers, and achieve a drastic improvements of resolution for more complicated objects. It is remarkable that the significant gain in resolution persists even for not sharp measurements, when there is a time-uncertainty in photon detecting, and the photons are registered in some intervals of delays.

Whereas the considered example of the antenna has rather model character, the described features of the sensing schemes have a general character. We believe that the temporal-spatial correlation structure exhibited by the field of the simplest  two-TLS quantum antenna, can be reproduced for other emitting systems. One can imagine, for example, that time-shaped photon pairs or high intensity two-mode squeezed pulses might are able to exhibit the required correlations \cite{brecht2015photon,ansari2018tailoring}. One can even try to approximate a necessary correlation structure by classically correlated antenna sources (as, for example, was demonstrated in recent work \cite{mikhalychev2018synthesis} on synthesis of quantum antennas). Use of high intensity source is also preferable taking into account background noise present in realistic far-field sensing scenarios.

We also demonstrated that measurements of the delayed intensity correlations can lead to super-resolution for other imaging schemes, such as near-field imaging, detecting the antenna position and inferring the distance between antenna emitters.
We believe that obtained results can have significant impact on the field of quantum imaging and far-field sensing.

D. M., A.P.N., A. M., and I. K.  acknowledge support from the EU projects Horizon-2020 SUPERTWIN id.686731 and PhoG id.820365, the National Academy of Sciences of Belarus program "Convergence", and the BRRFI project F18U-006.




\bibliography{Radar}



\end{document}